\documentclass{pasj00}

\DeclareAbbreviation\AAHam{Astron. Abh. Hamburg. Sternw.}
\DeclareAbbreviation\AARv{Astron. Astrophys. Rev.}
\DeclareAbbreviation\an{Astron. Nachr.}
\DeclareAbbreviation\AcA{Acta Astron.}
\DeclareAbbreviation\Afz{Astrofizika}
\DeclareAbbreviation\AnTok{Tokyo Astron. Obs. Annals, Sec. Ser.}
\DeclareAbbreviation\Ap{Astrophysics}
\DeclareAbbreviation\ARep{Astron. Rep.}
\DeclareAbbreviation\ATel{Astronomer's Telegram}
\DeclareAbbreviation\ATsir{Astron. Tsirk.}
\DeclareAbbreviation\AcApS{Acta Astrophys. Sinica}
\DeclareAbbreviation\AstL{Astron. Letters}
\DeclareAbbreviation\BaltA{Baltic Astron.}
\DeclareAbbreviation\BASI{Bull. Astron. Soc. India}
\DeclareAbbreviation\BeSN{Be Star Newsletter}
\DeclareAbbreviation\GCN{GCN}
\DeclareAbbreviation\ibvs{Inf. Bull. Variable Stars}
\DeclareAbbreviation\JAD{J. Astron. Data}
\DeclareAbbreviation\JAVSO{J. American Assoc. Variable Star Obs.}
\DeclareAbbreviation\JBAA{J. British Astron. Assoc.}
\DeclareAbbreviation\LowOB{Lowell Obs. Bull.}
\DeclareAbbreviation\MitVS{Mitteil. Ver\"{a}nderl. Sterne}
\DeclareAbbreviation\MmSAI{Mem. Soc. Astron. Ita.}
\DeclareAbbreviation\Msngr{Messenger}
\DeclareAbbreviation\NewA{New Astron.}
\DeclareAbbreviation\NewAR{New Astron. Rev.}
\DeclareAbbreviation\OAP{Odessa Astron. Publ.}
\DeclareAbbreviation\Obs{Observatory}
\DeclareAbbreviation\PASA{Publ. Astron. Soc. Australia}
\DeclareAbbreviation\PAZh{Pis'ma AZh}
\DeclareAbbreviation\PhR{Phys. Rep.}
\DeclareAbbreviation\PVSS{Publ. Variable Stars Sect. R. Astron. Soc. New Zealand}
\DeclareAbbreviation\PZ{Perem. Zvezdy}
\DeclareAbbreviation\PZP{Perem. Zvezdy Pril.}
\DeclareAbbreviation\QJRAS{QJRAS}
\DeclareAbbreviation\RMxAA{Rev. Mexicana Astron. Astrof.}
\DeclareAbbreviation\RvMA{Reviews of Modern Astron.}
\DeclareAbbreviation\Sci{Science}
\DeclareAbbreviation\SvA{Soviet Astronomy}
\DeclareAbbreviation\SvAL{Soviet Astronomy Letters}
\DeclareAbbreviation\VeSon{Ver\"{o}ff. Sternw. Sonneberg}
\DeclareAbbreviation\VSOLJBul{VSOLJ Variable Star Bull.}
\DeclareAbbreviation\yCat{VizieR Online Data Catalog}
\DeclareAbbreviation\ZA{Z. Astrophys.}

\def\ASPConf#1#2{ASP Conf. Ser. #1, #2}

\def\PublisherKluwer{Dordrecht: Kluwer Academic Publishers}
\def\PublisherASP{San Francisco: ASP}

\begin{document}
\SetRunningHead{M. Uemura, et al., }{WZ~Sge stars as the missing population}
\Received{2010/01/04}
\Accepted{2010/03/03}

\title{Dwarf Novae in the Shortest Orbital Period Regime:\\
II. WZ~Sge Stars as the Missing Population near the Period Minimum}

\author{
Makoto \textsc{Uemura}\altaffilmark{1},
Taichi \textsc{Kato}\altaffilmark{2},
Daisaku \textsc{Nogami}\altaffilmark{3}, and 
Takashi \textsc{Ohsugi}\altaffilmark{1}}

\altaffiltext{1}{Astrophysical Science Center, Hiroshima University, Kagamiyama
1-3-1, \\Higashi-Hiroshima 739-8526}
\email{uemuram@hiroshima-u.ac.jp}
\altaffiltext{2}{Department of Astronomy, Faculty of Science, Kyoto
  University, Sakyo-ku, Kyoto 606-8502}
\altaffiltext{3}{Kwasan Observatory, Kyoto University, Yamashina-ku,
  Kyoto 607-8471}

%


\maketitle

\begin{abstract}

WZ~Sge-type dwarf novae are characterized by long recurrence times of
outbursts ($\sim 10\;{\rm yr}$) and short orbital periods ($\lesssim
85\;{\rm min}$).  A significant part of WZ~Sge stars may remain
undiscovered because of low outburst activity.  Recently, the observed
orbital period distribution of cataclysmic variables (CVs) has changed
partly because outbursts of new WZ~Sge stars have been discovered
routinely.  Hence, the estimation of the intrinsic population of
WZ~Sge stars is important for the study of the population and
evolution of CVs.  In this paper, we present a Bayesian approach to
estimate the intrinsic period distribution of dwarf novae from
observed samples. In this Bayesian model, we assumed a simple
relationship between the recurrence time and the orbital period which
is consistent with observations of WZ~Sge stars and other dwarf novae.
As a result, the minimum orbital period was estimated to be $\sim
70\;{\rm min}$.  The population of WZ~Sge stars exhibited a spike-like
feature at the shortest period regime in the orbital period
distribution. These features are consistent with the orbital period
distribution previously predicted by population synthesis studies.  We
propose that WZ~Sge stars and CVs with a low mass-transfer rate are
excellent candidates for the missing population predicted by the
evolution theory of CVs.
\end{abstract}

\section{Introduction}

Cataclysmic variables (CVs) are close binary systems containing a
white dwarf and a Roche-lobe filling normal star.  They have several
subclasses, such as, dwarf novae (DNe), nova-like variables, novae, and
magnetic CVs (\cite{war95book}).  CVs are considered to be one of the
final stages of the evolution of low mass binaries.  The formation and
evolution of low mass binaries can be explored through the orbital
period ($P_{\rm orb}$) distribution of CVs (for a review, see
\cite{kin88binaryevolution}).
 
Stable mass-transfer processes in CVs are driven by angular momentum
removal from the binaries.  In short period CVs having $P_{\rm
  orb}\lesssim 3$~hr, the driving mechanism is considered to be
angular momentum removal associated with gravitational radiation
(\cite{pac81cvevolution}; \cite{rap82cvevolution}).  A CV evolves
toward a short-$P_{\rm orb}$ region by losing angular momentum.  This
phase is terminated when the core of the secondary star becomes
degenerate.  Then, the degeneracy pressure changes the mass--radius
relationship of the secondary star.  As a result, the CV
evolves toward a long-$P_{\rm orb}$ regime after passing through the
minimum $P_{\rm orb}$ ($P_{\rm min}$). 

The $P_{\rm orb}$ distribution near $P_{\rm min}$ has received 
attention, since the observed $P_{\rm orb}$ distribution has different
characteristics from that expected from theoretical studies.  It is
well known that $P_{\rm min}$ is $\sim 80$~min in the observed CV
population (\cite{pac81cvevolution}; \cite{rap82cvevolution}).
However, the theoretically predicted $P_{\rm min}$ is 60--70~min,
significantly shorter than the observed value
(e.g. \cite{kol93CVpopulation}).  This is the so-called 
``period minimum problem''.  
According to population synthesis studies, most CVs have already
passed through $P_{\rm min}$.  It has been proposed that the evolution
time-scale is 
long for an evolutionary-advanced system with a low mass 
secondary star, in other words, the mass-transfer rate ($\dot{M}$)
from the secondary star is low  (e.g. \cite{kol93CVpopulation}; 
\cite{how97periodminimum}).  As a result, a spike-like feature is
expected to appear near $P_{\rm min}$ in the $P_{\rm orb}$
distribution due to the accumulation of systems. However, the
observed distribution is rather flat.  This is the so-called
``period spike problem'' (\cite{kol99CVperiodminimum};
\cite{ren02CVminimum}). 

Thus, it is known that there is a ``missing'' population in the
observed CVs with a $P_{\rm orb}$ range of 70--80~min when compared with
the predicted population.  \citet{lit08postPmin} have recently
reported three CVs having a brown dwarf secondary, which are
candidates for the missing population (\cite{lit06sdss1035};
\cite{lit07sdss1507}).  This discovery indicates that CVs can actually
reach the post-$P_{\rm min}$ regime as predicted by the theories,
while the number of such systems is still not properly known.

It has been suggested that the problems can be reconciled by another
mechanism of angular momentum removal in addition to
gravitational radiation (\cite{rez01gr}), such as the effects of 
magnetic stellar wind braking (\cite{kin02CVperiodminimum}), the
circumbinary disk (\cite{wil05circumdisk}), and a combination of
magnetic propeller and accretion disk resonances (\cite{mat06mp}).

On the other hand, it is possible that the missing population exists,
but remains undiscovered.  According to the disk instability model for
DNe, the recurrence time of superoutbursts (supercycle; $T_s$) is longer
in a system with a lower $\dot{M}$ (\cite{osa95wzsge}).
WZ~Sge-type DNe are known to have long $T_s$ ($\gtrsim
10\;{\rm yr}$), which indicate a low $\dot{M}$
(\cite{how95TOAD}; \cite{kat01hvvir}).  It has been proposed that
several WZ~Sge stars, or CVs with a low $\dot{M}$ have remained
undiscovered because of their low activity (\cite{mey98wzsge}).  
Since WZ~Sge stars are found concentrating in a short $P_{\rm orb}$
regime, these objects are a candidate for the 
missing population (\cite{pat98evolution}).  However, their intrinsic
contribution to the whole CV population has been poorly understood
because their $T_s$ were uncertain and their quiescent luminosity is low. 
\citet{gan09sdssCV} have recently reported that the CV sample 
from the Sloan Digital Sky Survey (SDSS) database shows a period spike
feature in the $P_{\rm orb}$ range of 80--86~min.  The spike feature
is a consequence of deep images and a homogeneously selected CV sample
obtained by SDSS, which could detect intrinsically faint CVs with a
low $\dot{M}$. Recently, in addition, $T_s$ of several WZ~Sge stars
have been well determined owing to long monitoring of DNe by amateur
observers (e.g. \cite{waa07gwlib})

In this paper, we investigate the contribution of long-$T_s$ DNe
to the whole population of CVs by estimating their intrinsic
population.  In \textsection~2, we demonstrate
that the whole population of CVs has changed significantly due to a
recent increase of the number of short $P_{\rm orb}$ CVs.  In
\textsection~3, we estimate the population of DNe by considering the
detection probability of superoutbursts using Bayesian analysis.  We 
discuss the implications of our results to the study of CV
evolution in \textsection~5.  In the final section, we summarize our
findings. 

\section{Recent change in the observed $P_{\rm orb}$ distribution
  in CVs} 

\begin{figure}
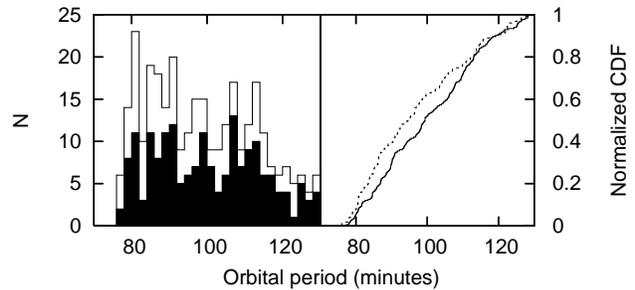

  \begin{center}
    \FigureFile(85mm,85mm){rkcat.eps}
  \end{center}
  \caption{$P_{\rm orb}$ distribution of CVs.  Left panel: 
  Sources in the 2003 sample (RKcat7.0, 
  black histogram) and 2008 sample (RKcat7.9, white 
  histogram).  Right panel: Normalized cumulative distribution functions 
  (CDF) of the 2003 (solid line) and the new samples discovered
  between 2003 and 2008 (dotted line).}\label{fig:rk}
\end{figure}

Both the period minimum and spike problems arise from the discrepancy
between the $P_{\rm orb}$ distribution predicted by the population
synthesis studies and the observed one.  In previous studies,
the predicted distribution was occasionally compared with the
distribution of all known CVs.  The left panel of figure~\ref{fig:rk}
shows the $P_{\rm orb}$ distribution of all known CVs with $P_{\rm
  orb}<130\;{\rm min}$ discovered before 2003 (black) and 2008
(white).  The samples were selected from Ritter \& Kolb catalog
\footnote{$\langle$http://www.mpa-garching.mpg.de/RKcat/$\rangle$}: 
184 sources in RKcat 7.0 for the 2003 sample and 312 sources in 
RKat 7.9 for the 2008 sample.  Both magnetic and non-magnetic CVs are
included in the samples.  As can be seen in the figure, the fraction
of short-$P_{\rm orb}$ 
systems in the 2008 sample is apparently larger than that in the 2003
sample.  We applied the Kolmogorov--Smirnov (KS) test to evaluate
whether the two samples differ significantly.  We calculated KS
probabilities, $P_{\rm KS}$, from the 2003 and 2008 samples, and also
from the 2003 sample and 128 new sources discovered between 2003 and
2008.  The results are listed in table~\ref{tab:kstest}.  The test for
the 2003 samples and the sample of new sources gives a low probability
of 0.07, and based on this we conclude that the distributions of the
samples differ significantly.  For the test with the full 2008 sample,
the probability is also low at 0.73 and hence it must be concluded
that the 2003 and 2008 samples differ overall. The difference between
the samples is also clear in the normalized cumulative distribution
functions (CDF) of the 2003 sample and the new sources in the 2008
sample, as shown in the right panel of figure~\ref{fig:rk}.  These
results demonstrate that the $P_{\rm orb}$ distribution of all known
CVs has significantly changed recently.  

\begin{table}
  \caption{Results of Kolmogorov--Smirnov tests between the 2003 
  (RKcat ver. 7.0) and 2008 (RKcat ver. 7.9) samples.}\label{tab:kstest}
  \begin{center}
    \begin{tabular}{cccc}
     \hline
     sample & $P_{\rm KS}^*$ & $N_1^\dag$ & $N_2^\dag$ \\
     \hline
     only new objects & 0.07 & 184 & 128 \\
     all objects      & 0.73 & 184 & 312 \\
     \hline
     \multicolumn{4}{l}{\footnotesize{$^*\; P_{\rm KS}$: Kolmogorov-Smirnov
     probability.}}\\
     \multicolumn{4}{l}{\footnotesize{$^\dag\; N_1$, $N_2$: the number of
     objects in the 2003 and 2008 samples.}}\\     
    \end{tabular}
  \end{center}
\end{table}

This change is mainly
caused by the outburst detection of new DNe and large spectroscopic
surveys.  First, outbursts of new and known WZ~Sge stars 
have routinely been discovered recently and hence the number of known
WZ~Sge stars has rapidly increased (\cite{sha07j0329};
\cite{waa07gwlib}; \cite{tem07v455and}; \cite{uem07j1021};
\cite{soe081600}; \cite{shu08j0238}; \cite{zha08j0804}).  The dramatic
increase of newly recognized SU~UMa-type and WZ~Sge-type DNe is also
apparent in tables in \citet{kat10pdot}, in which 14 WZ~Sge-type DNe
(out of a total of 32 systems; \cite{RKcat}; \cite{kat10pdot}) was
recognized after 2005. This increase of the outburst detection is
attributed to the development of automatic sky monitor systems
(e.g. \cite{ASAS3}; \cite{PioftheSky}; and most recently,
\cite{CRTS}), and improvements in the techniques and instruments of
amateur observers (notably by K. Itagaki).  The development of online
networks for transient phenomena has also played an important role in
circulating information on WZ~Sge stars, and enabled timely
observation essential for identifying their nature (\cite{VSNET}). 

Second, large spectroscopic surveys have also played an important
role in increasing the number of CVs recently, for example, SDSS and
the Hamburg Quasar Survey (HQS) (\cite{szk02SDSSCVs};
\cite{bad98RASSID}).  Those large surveys allow us to investigate the
$P_{\rm orb}$ distribution without complicated observation bias which
should be considered in a sample containing all CVs discovered by
various detection methods.  The $P_{\rm orb}$ distribution of all
known CVs would be same as before if the distribution of those new CVs
is same as that of the previously known CVs.  However, the
distribution of the SDSS CVs has a period spike feature, and is
different from the previously known distributions (\cite{gan09sdssCV}).

Thus, it is evident that the period minimum and spike problems should
be discussed not with the sample containing all known CVs, but with
the sample in which observation bias for the $P_{\rm orb}$
distribution can be simply evaluated. In the next section, we collect
outburst-selected DN samples, and estimate the intrinsic population of
DNe without the complicated bias for the $P_{\rm orb}$ distribution.

\section{Bayesian estimation of the intrinsic population of DNe}

As well as new WZ~Sge stars, outbursts of known WZ~Sge stars have
been observed routinely.  Thereby, we can now estimate $T_s$
more accurately for them.  In \citet{uem10j0557} (hereafter, Paper~I), we
revised $T_s$ of several DNe and found that WZ~Sge stars form a major
group in $P_{\rm orb}\lesssim 86\;{\rm min}$.  This implies that a DN
enters the WZ~Sge regime on an evolutionary course at $P_{\rm orb}\sim 
86\;{\rm min}$ as $\dot{M}$ decreases.  Using $T_s$ and $P_{\rm orb}$
from Paper~I, we consider a statistical experiment to estimate the
intrinsic $P_{\rm orb}$ distribution of DNe from observed systems
based on a simple model for $T_s$.  Using Bayesian
analysis, we calculate the posterior distributions of parameters
that characterize the $P_{\rm orb}$ distribution (e.g. \cite{tro08bayes}).  

\subsection{Overview of analysis}

We first present an outline of the analysis.  The models and samples are
detailed in subsequent sections.

We consider a DN sample, $\mathbf{x}=\{P_{{\rm orb},i}\}$, whose
superoutbursts were detected at least once in a certain period of
time.  The intrinsic $P_{\rm orb}$ population of DNe can be reproduced
from this outburst-selected sample if $T_s$ can be described with a
function of $P_{\rm orb}$.  The probability density
function (PDF) of observed DNe, $Q(P_{\rm orb})$, is given by the
intrinsic period distribution, $I(P_{\rm orb})$, and the selection
effects arising from the observation process, $D(P_{\rm orb})$, in the
form $Q=D\cdot I/A_Q$, where $A_Q$ is a normalization factor.  

With a parameter set, $\mathbf{y}$, that characterizes $I$, the
likelihood of $\mathbf{x}$, can be defined by 
\begin{eqnarray}
L(\mathbf{y}\mid \mathbf{x})=Q(\mathbf{x}\mid \mathbf{y})=\prod_i
Q(P_{{\rm orb},i}\mid \mathbf{y}).  
\end{eqnarray}
Posterior distributions of parameters, $P(\mathbf{y}\mid \mathbf{x})$,
are estimated with Bayes' theorem, as follows:
\begin{eqnarray}
P(\mathbf{y}\mid \mathbf{x})=
\frac{
  L(\mathbf{y}\mid \mathbf{x})P(\mathbf{y})}{
  \sum L(\mathbf{y}\mid \mathbf{x})P(\mathbf{y})}=
\frac{
  Q(\mathbf{x}\mid \mathbf{y})P(\mathbf{y})}{
  \sum Q(\mathbf{x}\mid \mathbf{y})P(\mathbf{y})}.
\end{eqnarray}
In our analysis, we used flat distributions for the prior
probabilities of parameters: $P(\mathbf{y})={\rm const}$.  

\subsection{Sample}

Samples were selected from SU~UMa-type DNe in RKcat7.9.  The ``ASAS''
sample consists of systems in which superoutbursts were recorded
at least once in the ASAS-3 database between 1 Jan 2003 and 31 Dec
2007 (\cite{ASAS3}).  A superoutburst is defined as an outburst
lasting for $\ge 5\;{\rm d}$ with a small fading rate ($\sim 0.1\;{\rm
  mag}\,{\rm d}^{-1}$).  $P_{\rm orb}$ of our sample was limited to
$\le 130\;{\rm min}$ because our simple model of $D$ cannot reproduce
the features of a period gap of 2~hr $\lesssim P_{\rm orb} \lesssim$
3~hr (for details, see the next section).  We
made another sample, called the ``mixed'' sample, using the procedure 
for data selection as for the ASAS
sample, but using both the VSNET and ASAS-3
databases for searching superoutbursts (\cite{VSNET}).

We excluded two minor groups of objects in the shortest $P_{\rm orb}$
regime.  The first group is ER~UMa stars 
(ER~UMa, IX~Dra, RZ~LMi, V1159~Ori, and DI~UMa), which are
characterized by atypically short $T_s$ (20--50~d) despite a short
$P_{\rm orb}$ (\cite{kat95eruma}).  The other group was identified in
Paper~I for the first time and called ``Group~X''.  The members of
this group have rather average $T_s$ ($\sim 400$~d) despite a short
$P_{\rm orb}$.  These two groups were excluded from our sample
because they may lead to an overestimate of the intrinsic population
of short-$P_{\rm orb}$ systems having long $T_s$.

\begin{figure}
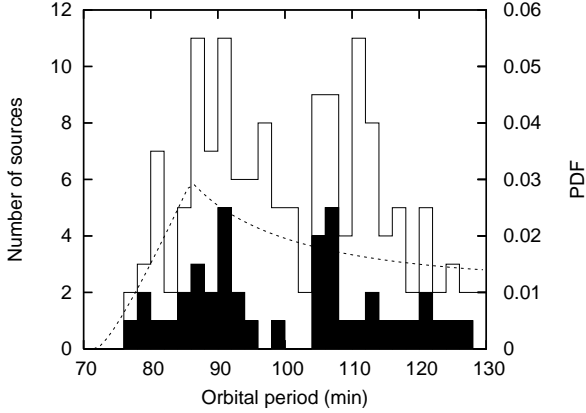

  \begin{center}
    \FigureFile(80mm,80mm){sample.eps}
  \end{center}
  \caption{Distribution of $P_{\rm orb}$ of the DN samples.
  The black and white histograms indicate the ASAS and mixed
  samples, respectively.  The dashed line indicates the best model PDF 
  derived from our analysis (for details, see the
  text).}\label{fig:sample}
\end{figure}

We obtained 42 and 146 sources for the ASAS and mixed
samples.  The ASAS sample is expected to be an ideal sample without
complicated selection effects since the superoutbursts were detected
through a single observation process defined by the ASAS system.  The
mixed sample may be affected by complicated selection effects.  We
calculated the KS probability of both samples.  The obtained
probability is unexpectedly high, $0.98$, which means that there is no
significant difference between those two samples.  This implies that our
method of data selection has the advantage of reducing selection
bias.  The $P_{\rm orb}$ distributions of these two samples are shown
in figure~\ref{fig:sample}.

\subsection{Model of the $P_{\rm orb}$ distribution}

The intrinsic $P_{\rm orb}$ distribution, $I$, must be
characterized by at least two parameters, namely, one for the degree
of accumulation near $P_{\rm min}$ and the other for the real 
$P_{\rm min}$.  We express $I$ by the following formulae: 
\begin{eqnarray}
I(p)= \left \{
\begin{array}{ll}
p^{-\alpha} e^{-\alpha/p}/A_I & (p\ge 1) \\
0 & (p< 1)
\end{array} \right. \\
p = P_{\rm orb}-P_{\rm min}\;({\rm min}),
\end{eqnarray}
where $\alpha$ is a parameter for the degree of the accumulation
near $P_{\rm min}$.  These are the parameters of our Bayes model,
that is, $\mathbf{y}=\{\alpha, P_{\rm min}\}$.  
A large $\alpha$ yields a distribution with a prominent spike
near $P_{\rm min}$.  The distribution becomes flat in the case of 
$\alpha=0$.  For a negative $\alpha$, the distribution has a peak in the
long-$P_{\rm orb}$ region.  The function $I$ has an extremum at 
$P_{\rm orb}=P_{\rm min}+1$.  $A_I$ is a normalization factor.  
While this formula of $I$ itself has no physical meaning, it can
approximate the profile of the $P_{\rm orb}$ distribution below the
period gap predicted by population synthesis studies, with high
$\alpha$ ($\alpha\sim 1$) (e.g. \cite{bar03CVevolv}).  This model
should be used only below the period gap ($2\;{\rm hr}\lesssim P_{\rm
  orb} \lesssim 3\;{\rm hr}$; \cite{spr83periodgap}), since it cannot
reproduce the gap structure. 

We considered a simple model for the selection effect, $D$.  First, we
considered that all the systems are pre-$P_{\rm min}$ CVs and neglected
contributions from post-$P_{\rm min}$ CVs.  The contribution of
post-$P_{\rm min}$ CVs to our analysis is discussed in
\textsection~4.2.  Second, we used a proposed model for the relationship
between $T_s$ and $P_{\rm orb}$ based on observations.  In Paper~I, we
showed that most systems are ordinary SU~UMa stars in a $P_{\rm
  orb}$ range of $\gtrsim 86\;{\rm min}$.  The average $T_s$ is 470~d
in $86\;{\rm min} \leq P_{\rm orb}\leq 95\;{\rm min}$.  On the other
hand, in the shortest $P_{\rm orb}$ regime of $\lesssim 86\;{\rm
  min}$, the major group is WZ~Sge stars.  Based on these results in
Paper~I, we assumed the following for our model, $D$: i) $T_s$ is
constant at $470\;{\rm d}$ in $P_{\rm orb}\ge P_{\rm crit}$.  ii)
$T_s$ becomes longer with shorter $P_{\rm orb}$ in $P_{\rm orb}<
P_{\rm crit}$.  iii) We set $P_{\rm crit}=86\;{\rm min}$.  

\begin{figure}
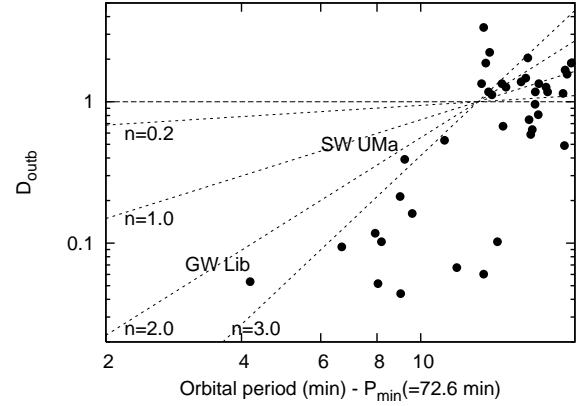

  \begin{center}
    \FigureFile(80mm,80mm){doutb.eps}
  \end{center}
  \caption{Outburst detection probability, $D_{\rm outb}$, as a
  function of $P_{\rm orb}-P_{\rm min}$.  $P_{\rm min}$ is assumed to
  be 72.6~min in the figure (for more detail, see the text).  The
  filled circles indicate SU~UMa stars whose $T_s$ is known.  The
  dashed lines indicate the models with various $n$.  Two objects,
  GW~Lib and SW~UMa are labeled.  They are used to estimate a lower
  limit of $n$.}\label{fig:doutb}   
\end{figure}

The outburst detection probability, $D_{\rm outb}$, is inversely
proportional to $T_s$.  Taking $D_{\rm outb}$ at $T_s=470\;{\rm d}$ to
be unity, the above assumptions can be expressed as follows: 
\begin{eqnarray}
D_{\rm outb}(P_{\rm orb})= \left \{
\begin{array}{ll}
1 & (P_{\rm orb}\ge P_{\rm crit})\\
\left( \frac{P_{\rm orb}-P_{\rm min}}{P_{\rm crit}-P_{\rm min}}
\right)^n & (P_{\rm orb}< P_{\rm crit}) 
\end{array} \right.
\end{eqnarray}
The parameter $n$ can be estimated from the observed $T_s$ listed in 
table~5 of Paper~I.  Using GW Lib ($P_{\rm orb}=76.7808$~min and
$T_s=8800$~d) and SW UMa ($P_{\rm orb}=81.8136$~min and $T_s=1200$~d),
for example, we obtained $n=2.5$ and $P_{\rm min}=72.6\;{\rm min}$. 
We show $D_{\rm outb}$ calculated from the observed $T_s$ of our
sample with $P_{\rm min}=72.6\;{\rm min}$ in figure~\ref{fig:doutb}.
The dashed lines in the figure denote the models, $D_{\rm outb}$, with
various $n$.  This figure shows that the estimation of the parameters
with GW~Lib and SW~UMa yields a lower-limit of $n$ for our sample.  
As discussed in \S~4.1, a larger $n$ leads to a larger number of 
short-$P_{\rm orb}$ DNe.  In the following analysis, we used $n=2.0$, which
gives a firm lower-limit for $n$, as can be seen from
figure~\ref{fig:doutb}.  $n$ can be smaller than 2.0 only when $P_{\rm
  min}> 72.6\;{\rm min}$.  We check whether our estimation of $n$ is
consistent with the results of the analysis in \S~4.1.

Finally, we considered the relationship between the absolute
magnitude, $M_V$, at supermaximum and $P_{\rm orb}$.  Since the ASAS
sample is a magnitude-limited sample, the selection bias caused by the
$P_{\rm orb}$ dependence of $M_V$ should be corrected.  Under an
assumption of a uniform space distribution of DNe, the number of
detectable systems, $D_{\rm mag}$, can be written as a function of
$M_V$ as 
\begin{eqnarray}
D_{\rm mag}(P_{\rm orb})\propto 10^{-0.6 M_V(P_{\rm orb})}.
\end{eqnarray}
\citet{har04MVPorb} reported the relationship between $M_V$ at normal
outburst maxima and $P_{\rm orb}$.  In addition to their relationship,
\citet{pre07evolution} proposed that WZ~Sge stars have a constant
$M_V$, independent of $P_{\rm orb}$.  We follow their models for $M_V$
at the supermaxima of ordinary SU~UMa stars ($M_{V,{\rm SU}}$) and WZ~Sge
stars ($M_{V,{\rm WZ}}$), that is: 
\begin{eqnarray}
\begin{array}{ll}
M_{V,{\rm SU}}(P_{\rm orb})=C_1 - 0.383 P_{\rm orb}&
(P_{\rm orb}\ge P_{\rm crit})\\
M_{V,{\rm WZ}}(P_{\rm orb})=C_2 & (P_{\rm orb}< P_{\rm crit}) 
\end{array},
\end{eqnarray}
where $C_2$ is a constant determined by $M_{V,{\rm SU}}
(P_{\rm crit})=M_{V,{\rm WZ}}(P_{\rm crit})$.  The unit of $P_{\rm
  orb}$ is hour.  The constant $C_1$ is not important in our
analysis. 

Finally, the model of the $P_{\rm orb}$ distribution of DNe is
expressed by the PDF, $Q$, using the two parameters as follows: 
\begin{eqnarray}
Q(a,P_{\rm min})=
D_{\rm mag}\cdot D_{\rm outb}\cdot I(a,P_{\rm min})/A_Q(a,P_{\rm min}). 
\end{eqnarray}

\subsection{Bayesian analysis and results}

We calculated the likelihood of the sample using equation~(1).  We
produced posterior probability distributions of $\alpha$ and $P_{\rm
  min}$ using the Markov Chain Monte Carlo method (MCMC)
(\cite{met53mcmc}).  The 
calculation procedure of MCMC is as follows: At the $n$-th step of
MCMC, we obtain the likelihood, $L_n=L(a_n,P_{{\rm min},n}\mid
\mathbf{x})$.  We then randomly move to another point in the parameter
space of $\alpha$ and $P_{\rm min}$ by adding random values drawn from
a Gaussian distribution which has zero mean and dispersion chosen to
efficiently sample the likelihood surface.  As a result, we obtain
$L_{n+1}$.  We count the parameters of the $(n+1)$-th step if
$L_{n+1}>L_n$ or $L_{n+1}/L_n$ is larger than a uniform random number
between 0 and 1. Otherwise, the $n$-th parameters are substituted for 
the $(n+1)$-th ones.  After discarding the first 100 steps, we sample
every 100 steps until the size of the sample reaches $10^5$.  This
cycle forms a set of samples for $\alpha$ and $P_{\rm min}$. We
obtained ten sets of samples with different initial values of $\alpha$
and $P_{\rm min}$. We confirmed that the obtained ten samples had the
same $P_{\rm orb}$ distribution.  We merged them and finally obtained
the posterior distributions.

\begin{figure}
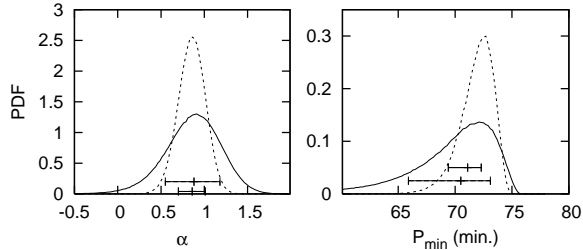

  \begin{center}
    \FigureFile(80mm,80mm){post.eps}
  \end{center}
  \caption{Posterior probability distributions of $\alpha$ and $P_{\rm
  min}$.  The solid and dotted lines were calculated from the ASAS and
  mixed samples, respectively.  The medians and 68.3~\% confidence
  levels of the parameters are also indicated.}\label{fig:post}
\end{figure}

\begin{figure}
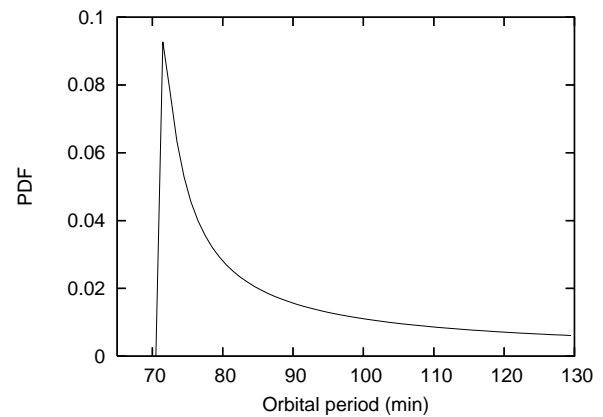

  \begin{center}
    \FigureFile(80mm,80mm){model.eps}
  \end{center}
  \caption{Intrinsic $P_{\rm orb}$ distribution of DNe, $I$, estimated
  from the observed sample.}\label{fig:model}
\end{figure}

Figure~\ref{fig:post} shows the posterior distributions of $\alpha$
(left panel) and $P_{\rm min}$ (right panel).  Each panel contains both 
the results estimated from the ASAS and mixed samples.  The median and
68.3\% confidence level of $\alpha$ and $P_{\rm min}$ are summarized
in table~\ref{tab:post} and are also indicated in the figure.  As
expected from the KS test in \S~3.2, the parameters estimated from the
ASAS sample agree with those from the mixed sample within the errors.
The best model of $Q$ for the ASAS sample is indicated by the dashed
line in figure~\ref{fig:sample}.  The intrinsic $P_{\rm orb}$
distribution, $I$, is shown in figure~\ref{fig:model}.

As shown in table~\ref{tab:post}, $\alpha$ was estimated to be high,
namely, the intrinsic $P_{\rm orb}$ distribution, $I$, has a
spike near $P_{\rm min}$, as can be seen in figure~\ref{fig:model}.  
Short-$P_{\rm orb}$ systems in $P_{\rm orb}<P_{\rm crit}(=86\;{\rm
  min})$ account for 59~\% 
of the total number of samples between $P_{\rm min}<P_{\rm
  orb}<130$~min.  The estimated $P_{\rm min}$ is about 70~min,
significantly shorter than the observed $P_{\rm min}$ of about 78~min.

\begin{table}
  \caption{Best parameters derived from Bayesian analysis.}\label{tab:post}
  \begin{center}
    \begin{tabular}{cccccc}
     \hline 
     Sample & $n$ & $P_{\rm crit}$ & $M_V$ & $\alpha$   & $P_{\rm min}$ \\
            &     & (min)          &       &       & (min) \\
     \hline
     ASAS   & 2.0 & 86.0 & eq.(7) & $0.88^{+0.30}_{-0.33}$ & $70.5^{+2.6}_{-4.6}$ \\
     all RKcat  & 2.0 & 86.0 & eq.(7) & $0.86^{+0.15}_{-0.16}$ & $71.1^{+1.2}_{-1.7}$ \\
     \hline
     ASAS   & 1.0 & 86.0 & eq.(7) & $0.61^{+0.29}_{-0.32}$ & $71.3^{+1.9}_{-4.0}$ \\
     ASAS   & 3.0 & 86.0 & eq.(7) & $1.07^{+0.34}_{-0.37}$ & $66.2^{+3.8}_{-6.3}$ \\
     ASAS   & 2.0 & 82.0 & eq.(7) & $0.60^{+0.27}_{-0.32}$ & $71.3^{+2.5}_{-4.5}$ \\
     ASAS   & 2.0 & 90.0 & eq.(7) & $1.15^{+0.32}_{-0.36}$ & $69.5^{+2.6}_{-4.7}$ \\
     ASAS   & 2.0 & 86.0 & const. & $0.64^{+0.33}_{-0.38}$ & $68.6^{+3.0}_{-5.3}$ \\
     ASAS   & 0.2 & 86.0 & const. & $0.10^{+0.25}_{-0.28}$ & $73.0^{+1.2}_{-2.7}$ \\
     \hline
    \end{tabular}
  \end{center}
\end{table}

The estimated $P_{\rm orb}$ distribution, $I$, is analogous to those
expected from population synthesis studies
(e.g. \cite{kol93CVpopulation}), in terms of the accumulation of
systems near $P_{\rm min}$ and the short $P_{\rm min}$ of $\sim
70\;{\rm min}$.  We suggest that i) DNe having long $T_s$ are an
excellent candidate for the missing population near $P_{\rm min}$, and
ii) the observed $P_{\rm min}$ is significantly longer than the real
$P_{\rm min}$ because of the low outburst activity of DNe near $P_{\rm
  min}$. 

\section{Discussion}

\subsection{Dependence of $\alpha$ and $P_{\rm min}$ on the assumed
  model parameters} 

In the previous section, we assumed several parameters in our model to
estimate $\alpha$ and $P_{\rm min}$, such as $n$, $P_{\rm crit}$ and
$M_V$ at supermaximum.  Here, we discuss the dependence of the result
on these parameters.  We present $\alpha$ and $P_{\rm min}$ estimated
with several different sets of these parameters in table~\ref{tab:post}.

As can be seen in table~\ref{tab:post}, $\alpha$ is sensitive to both
$n$ and $P_{\rm crit}$.  Smaller $n$ or $P_{\rm crit}$ leads to a
smaller $\alpha$.  $\alpha$ also depends on the model of $M_V$.  It is
possible that $M_V$ takes a form different from equation~(7).
According to \citet{har04MVPorb}, $M_{V,{\rm SU}}$ for
superoutbursts has a large dispersion and does not
strictly follow the $M_V$--$P_{\rm orb}$ relationship for normal
outbursts.  This implies that the $P_{\rm orb}$ dependence of $M_{V,{\rm
SU}}$ is weak.  Assuming $M_{V,{\rm SU}}=M_{V,{\rm WZ}}={\rm const.}$,
we obtained a smaller $\alpha$ as shown in table~\ref{tab:post}.  
In all the above cases, $\alpha$ takes high values of $\gtrsim 0.6$
and a spike near $P_{\rm min}$ appears in the $P_{\rm
orb}$ distribution.  The spike feature disappears only when $n$ is
extremely small.  For example, we show the case of $n=0.2$ in
table~\ref{tab:post}.  Such a small $n$ is, however, unfavorable for the
observed $D_{\rm outb}$--$P_{\rm orb}$ relationship, as shown in
figure~\ref{fig:doutb}.

Table~\ref{tab:post} indicates that $P_{\rm min}$ is sensitive to $n$;
a small $n$ yields a large $P_{\rm min}$.  On the other hand, the
estimated $P_{\rm min}$ in table~2 are significantly smaller than the
observed value, even in the extreme case of $n=0.2$.  Thus, the result
is robust within the reasonable ranges of the model parameters.  

As mentioned in \textsection~3.3, we used $n=2.0$, which is a lower
limit of $n$ in the case of $P_{\rm min}=72.6\;{\rm min}$.  $n$ can be
smaller than 2.0 only when $P_{\rm  min}> 72.6\;{\rm min}$.  As shown
in table~2, all $P_{\rm min}$ satisfies $P_{\rm min}<72.6\;{\rm min}$,
except for the extreme case of $n=0.2$.  Therefore, it is less
likely that $n$ is smaller than 2.0 in our model. 

Our result highly depends on the assumption for
the $D_{\rm outb}$--$P_{\rm orb}$ relationship.  The assumption might
be incorrect if a number of short-$T_s$ systems remain undiscovered in
the shortest $P_{\rm orb}$ regime, such as {\it Group X} identified in
Paper~I.  The assumption for the $D_{\rm outb}$--$P_{\rm
orb}$ relationship should thus be reconsidered if the intrinsic population
of those objects is found to be much larger than the currently observed population.
However, it is unlikely that there are significant undetected systems because the outburst detection probability
must be high in such short-$T_s$ systems and hence their outbursts
should have already been discovered.

\subsection{Implications for CV evolution}

In this section, we discuss the implications of our result to CV
evolution, considering the presence of post-$P_{\rm min}$ CVs,
magnetic CVs, and past theoretical and observational works in this
field.

In the previous section, we regarded all systems as pre-$P_{\rm min}$
CVs.  However, it has been proposed that about $70\%$ of CVs have
already passed through $P_{\rm min}$ (\cite{kol93CVpopulation}; 
\cite{how97periodminimum}).  It is posible that all known DNe are
pre-$P_{\rm min}$ CVs, while most CVs have already passed through
$P_{\rm min}$ and remained undiscovered.  In this case, the 
intrinsic $P_{\rm orb}$ distribution would have a taller spike near
$P_{\rm min}$ than that in figure~\ref{fig:model} because of the
contribution of the undiscovered post-$P_{\rm min}$ CVs. 

On the other hand, a fraction of the long-$T_s$ DNe in our sample might
actually be post-$P_{\rm min}$ systems because post-$P_{\rm min}$
systems presumably have quite low $\dot{M}$ and long $T_s$.  Even
if most known WZ~Sge stars are post-$P_{\rm min}$ CVs, the spike
structure in figure~\ref{fig:model} is still expected to appear
because the spike feature depends, in principle, on the number of
long-$T_s$ DNe 
near $P_{\rm min}$ and is independent of the nature of the systems, namely,
whether they are pre- or post-$P_{\rm min}$ systems.  

In contrast to the spike feature, the real $P_{\rm min}$ is possibly
sensitive to the number of pre-$P_{\rm min}$ systems in the shortest
$P_{\rm orb}$ regime.  If the true fraction of post-$P_{\rm min}$ systems
is large in known WZ~Sge stars, then it means that superoutbursts of the
post-$P_{\rm min}$ systems are detectable for us.  If this is the
case, we should be able to detect outbursts of DNe below the observed
$P_{\rm min}$.  The estimated $P_{\rm min}$ could then be close to
the observed value. 

The true intrinsic fraction of post-$P_{\rm min}$ systems in the
shortest $P_{\rm orb}$ regime is poorly known, while recent
observational studies suggest the presence of post-$P_{\rm min}$
systems. \citet{two09mdot} investigated the temperature of the white
dwarf in CVs as a probe of $\dot{M}$.  Their result shows that there
are anomalies having quite low temperatures of the white dwarf
compared with most CVs in the shortest $P_{\rm orb}$ regime.  The
$\dot{M}$ of those anomalies are possibly lower than the ordinary
ones, and hence, suggesting that they are post-$P_{\rm min}$ systems.
\citet{lit08postPmin} reported three CVs in which the post-$P_{\rm
  min}$ nature was dynamically confirmed (also see,
\cite{lit06sdss1035}; \cite{lit07sdss1507}).  Among known DNe, WZ~Sge
has the longest $T_s$ and hence is a good candidate for post-$P_{\rm
  min}$ systems. \citet{ste07wzsge}, however, reported that its
secondary mass is so high that WZ~Sge is in fact a pre-$P_{\rm min}$
system.  The nature of WZ~Sge implies that most DNe whose $T_s$ are
known are still in the pre-$P_{\rm min}$ regime. 

As mentioned in \textsection~3.3, our model of $D_{\rm outb}$ assumed
a decrease of $\dot{M}$ toward $P_{\rm min}$ in the region $P_{\rm
  orb}<86\;{\rm min}$.  This was required from the observed
$T_s$--$P_{\rm orb}$ 
relationship, as can be seen in figure~\ref{fig:doutb}.  However, this is
apparently  inconsistent with theoretical calculations for the CV
evolution, which predict a rather constant $\dot{M}$ for 
pre-$P_{\rm min}$ systems even near $P_{\rm min}$
(e.g. \cite{sha92DNperiod}; \cite{kol93CVpopulation};
\cite{bar03CVevolv}).  If the theoretical prediction is true, we would
expect many short-$T_s$ systems in the shortest $P_{\rm
orb}$ regime.  The prediction is, however, unfavorable for the
observed population, because there are only a few short-$T_s$ systems in
$P_{\rm orb}\lesssim 86\;{\rm min}$, such as {\it Group X} in
Paper~I.  This inconsistency implies that the current understanding of
the secondary star is imperfect, since the theoretical
$\dot{M}$--$P_{\rm orb}$ relationship depends on the structure of the
secondary star.  \citet{mey99diskviscosity} proposed that the magnetic
activity of the secondary star reduces rapidly as it becomes
degenerate, and therby, the viscosity in the disk reduces and $T_s$ 
increases.  The observed $T_s$--$P_{\rm orb}$
relationship may be reprocuded if $T_s$ increases by this mechanism in 
$P_{\rm orb}<86\;{\rm min}$.

It has been reported that the $P_{\rm orb}$ distribution of magnetic
CVs is similar to that of non-magnetic CVs (e.g. \cite{wil05circumdisk}).
However, this discussion was based on samples obtained from various
observation processes, such as samples in RKcat.  As mentioned in
section~2, such a sample is not suitable to be compared with the
theoretical $P_{\rm orb}$ distribution obtained from population 
systhesis studies.  Our result suggests that, in the case of DNe, the
period minimum and spike problems are reconciled with a number of
undiscovered WZ~Sge stars.  Since AM~Her stars have no accretion disk,
and thereby, experience no DN-type outbursts, our result also implies
that another mechanism to reduce short-$P_{\rm orb}$ systems should
work for magnetic CVs (e.g. \cite{kol95magCV};
\cite{mey99amherevolution}).  The reduce of magnetic braking in 
magnetic CVs may lead to a different $P_{\rm orb}$ distribution from
non-magnetic CVs (\cite{web02polarperiodgap}; \cite{ara05mCV}).

\citet{gan09sdssCV} have recently reported the discovery of the period
spike feature in the CV sample selected from the SDSS database.  The
appearance of the spike is in agreement with our result; we reproduced
a spike based on observed outburst activity, while \citet{gan09sdssCV}
obtained it directly from the deep survey.  They discovered that CVs
having a white-dwarf dominated spectrum form a major population in the
CVs making the period spike feature.  Their spectrum indicates that
$\dot{M}$ of those CVs are quite low.  Hence, they may be WZ~Sge
stars, and their outbursts could be observed with a long $T_s$.
Actually, SDSS~J080434.20+510349.2, a system having a white-dwarf
dominated spectrum, experienced a WZ~Sge-type superoutburst in 2005
(\cite{pav07j0804}).  The SDSS CVs having a white-dwarf dominated
spectrum probably have the same nature as the objects that our
analysis predicts to be present in the shortest $P_{\rm orb}$ regime. 

On the other hand, the implications for the period minimum problem are
different between our result and \citet{gan09sdssCV}.  In our result,
the observed $P_{\rm min}$ is significantly 
longer than the real $P_{\rm min}$ because $T_s$ is so long that the
detection frequency is low near the real $P_{\rm min}$.  The SDSS
images are so deep that we can expect to detect CVs with a low
$\dot{M}$ near the real $P_{\rm min}$.  However, a sharp cut-off
exists at $P_{\rm orb}\sim 80\;{\rm min}$ in the $P_{\rm orb}$
distribution of the SDSS sample and only a few sources were found in
the $P_{\rm orb}$ range of 60--78~min.  The period minimum problem,
hence, remains unresolved in \citet{gan09sdssCV}.  The lack of the
sources in the $P_{\rm orb}$ range 60--78~min may indicate that a
fraction of known WZ~Sge stars are post-$P_{\rm min}$ CVs, as
discussed above.  Alternatively, some CVs with a very low $\dot{M}$
may remain undiscovered even in the SDSS survey.  \citet{gan09sdssCV}
identified CVs based on spectral features of objects that were
selected by several color criteria.  A part of the color criteria
plays a role in excluding single white dwarfs from the CV sample.  The
criteria, however, possibly excludes evolved CVs which have a quite
low $\dot{M}$, and thereby, have optical spectra dominated by the
emission from a white dwarf.

\section{Summary}

We estimated the intrinsic $P_{\rm orb}$ distribution of DNe assuming
a simple relationship between $T_s$ and $P_{\rm orb}$.  The estimated 
distribution has a spike near $P_{\rm min}$ and a shorter
$P_{\rm min}$ than the observed value, which was expected by the
theory of binary evolution.  We suggest that the nature of the
``missing'' population is DNe with long $T_s$ near $P_{\rm min}$.  

We thank B. G\"{a}nsicke for his valuable comments and
suggestions on this paper. 
This work was partly supported by a Grand-in-Aid from the Ministry of
Education, Culture, Sports, Science, and Technology of Japan
(19740104).

\appendix
\section{Outburst-selected DNe sample}

Table 3 and 4 lists the objects of the ASAS and mixed sample used in
our analysis described in section 3, respectively.  The object name
and $P_{\rm orb}$ were quoated from Ritter~\& Kolb catalog 7.9
(\cite{RKcat}). 
 
\begin{table*}
  \caption{Members of the ASAS sample.}\label{tab:asas}
  \begin{center}
    \begin{tabular}{rr|rr|rr|rr}
     \hline
     Object & $P_{\rm orb}$ (min) & Object & $P_{\rm orb}$ (min) &
     Object & $P_{\rm orb}$ (min) & Object & $P_{\rm orb}$ (min) \\
     \hline
     GW Lib       & 76.8 &
     J0233$-$1047 & 78.9 &
     J0137$-$0912 & 79.7 &
     V1108 Her    & 81.9 \\
     WX Cet       & 83.9 &
     J1112$-$3538 & 84.1 &
     CC Scl       & 84.5 &
     2219$+$1824  & 86.3 \\ 
     V1040 Cen    & 86.8 &
     AQ Eri       & 87.8 &
     J0918$-$2942 & 88.8 &
     J1025$-$1542 & 89.3 \\
     V436 Cen     & 90.0 & 
     EK TrA       & 90.5 &
     OY Car       & 90.9 &
     J1600$-$4846 & 91.3 \\
     J1536$-$0839 & 91.6 &
     J0232$-$3717 & 93.2 &
     AK Cnc       & 93.7 &
     GO Com       & 94.8 \\
     RZ Sge       & 98.3 &
     DT Oct       &104.5 &
     V368 Pe      &105.1 &
     VZ Pyx       &105.6 \\
     CC Cnc       &105.9 &
     J1556$-$0009 &106.7 &
     VW Hyi       &107.0 &
     Z Cha        &107.3 \\
     QW Ser       &107.4 &
     WX Hyi       &107.7 &
     RZ Leo       &109.5 &
     DH Aql       &111.4 \\
     J0219$-$3045 &112.9 &
     CU Vel       &113.0 &
     J0549$-$4921 &115.5 &
     V877 Ara     &116.7 \\
     TU Crt       &118.2 &
     TY PsA       &121.1 &
     KK Tel       &121.7 &
     V364 Peg     &122.4 \\
     YZ Cnc       &125.0 &
     GZ Cnc       &126.9 \\
     \hline
    \end{tabular}
  \end{center}
\end{table*}

\begin{table*}
  \caption{Members of the mixed sample.}\label{tab:asas}
  \begin{center}
    \begin{tabular}{rr|rr|rr|rr}
     \hline
     Object & $P_{\rm orb}$ (min) & Object & $P_{\rm orb}$ (min) &
     Object & $P_{\rm orb}$ (min) & Object & $P_{\rm orb}$ (min) \\
     \hline
     J0329$+$1250 & 76.0 &
     GW Lib       & 76.8 &
     J0222$+$4122 & 78.9 & 
     J0233$-$1047 & 78.9 \\
     J0137$-$0912 & 79.7 &
     SS LMi       & 80.2 &
     J1021$+$2349 & 80.5 &
     J0025$+$1217 & 80.7 \\
     V455 And     & 81.1 &
     AL Com       & 81.6 &
     SW UMa       & 81.8 &
     J1839$+$2604 & 81.9 \\
     J1959$+$2242 & 83.5 &
     WX Cet       & 83.9 &
     J1112$-$3538 & 84.1 &
     CC Scl       & 84.5 \\
     FL TrA       & 84.8 &
     J0804$+$5103 & 85.0 &
     V585 Lyr     & 85.4 &
     2219$+$1824  & 86.3 \\
     J0807$+$1138 & 86.4 &
     CI UMa       & 86.4 &
     DV Dra       & 86.4 &
     RX Vol       & 86.5 \\
     MM Sco       & 86.7 &
     V1040 Cen    & 86.8 &
     V1028 Cyg    & 86.9 &
     UZ Boo       & 87.4 \\
     AQ Eri       & 87.8 &
     V1454 Cyg    & 87.8 &
     XZ Eri       & 88.1 & 
     V4140 Sgr    & 88.5 \\
     J0918$-$2942 & 88.8 &
     J1025$-$1542 & 89.3 &
     V1141 Aql    & 89.3 &
     V402 And     & 89.4 \\
     V2051 Oph    & 89.9 &
     V436 Cen     & 90.0 & 
     BC UMa       & 90.2 &
     HO Del       & 90.3 \\
     EK TrA       & 90.5 &
     TV Crv       & 90.6 &
     J1227$+$5139 & 90.7 &
     OY Car       & 90.9 \\
     J1600$-$4846 & 91.3 &
     J1536$-$0839 & 91.6 &
     MR UMa       & 91.6 &
     Var 79 Peg   & 91.7 \\
     DO Vul       & 92.2 &
     J1653$+$2010 & 92.7 &
     J0232$-$3717 & 93.2 &
     UV Per       & 93.5 \\
     CT Hya       & 93.5 &
     AK Cnc       & 93.7 &
     AO Oct       & 94.1 &
     DM Lyr       & 94.2 \\   
     GO Com       & 94.8 &
     V551 Sgr     & 94.9 &
     BC Dor       & 95.2 &
     FQ Mon       & 95.8 \\
     SX LMi       & 96.8 &
     V701 Tau     & 96.9 &
     SS UMi       & 97.6 &
     NSV 4838     & 97.6 \\
     BZ UMa       & 97.9 &
     J0824$+$4931 & 97.9 &
     KS UMa       & 97.9 &
     V337 Cyg     & 97.9 \\
     RZ Sge       & 98.3 &
     TY Psc       & 98.4 &
     IR Gem       & 98.5 &
     V699 Oph     & 98.6 \\
     V1208 Tau    & 99.1 & 
     V1504 Cyg    &100.1 &
     CY UMa       &100.2 &
     RU Hor       &100.4 \\
     BB Ari       &101.2 &
     VW CrB       &101.8 & 
     PU Per       &102.7 &
     FO And       &103.1 \\
     AW Sge       &104.3 &
     TU Tri       &104.3 &
     DT Oct       &104.5 &
     OU Vir       &104.7 \\
     V1212 Tau    &105.1 &
     V368 Peg     &105.1 &
     KV And       &105.6 &
     VZ Pyx       &105.9 \\
     CC Cnc       &105.9 & 
     AY Lyr       &106.1 &
     IY UMa       &106.4 &
     J1556$-$0009 &106.7 \\
     VW Hyi       &107.0 &
     Z Cha        &107.3 &
     QW Ser       &107.4 &
     J0242$-$2802 &107.4 \\
     BE Oct       &107.6 &
     WX Hyi       &107.7 & 
     RZ Leo       &109.5 &
     QY Per       &109.6 \\
     0417$+$7445  &109.9 &
     SU UMa       &109.9 & 
     V630 Cyg     &110.0 &
     J1730$+$6247 &110.2 \\
     V1113 Cyg    &110.3 &
     AW Gem       &110.6 &
     HS Vir       &110.7 &
     EG Aqr       &110.9 \\
     TT Boo       &110.9 &
     V344 Pav     &110.9 &
     DH Aql       &111.4 &
     TY Vul       &111.9 \\
     V503 Cyg     &111.9 & 
     PV Per       &112.0 &
     V359 Cen     &112.2 &
     V660 Her     &112.7 \\
     BZ Cir       &112.9 &
     J0219$-$3045 &112.9 &
     CU Vel       &113.0 &
     J2100$+$0044 &113.8 \\
     NSV 14652    &113.8 & 
     BR Lup       &114.5 &
     NSV 9923     &114.6 &
     J2258$-$0949 &115.2 \\
     J0549$-$4921 &115.5 &
     RX Cha       &116.6 &
     V877 Ara     &116.7 &
     AB Nor       &117.1 \\
     CP Dra       &117.5 &
     V369 Peg     &117.6 & 
     TU Crt       &118.2 &
     HV Aur       &118.5 \\
     EF Peg       &120.5 &
     TY PsA       &121.1 &
     BF Ara       &121.7 &
     KK Tel       &121.7 \\
     V452 Cas     &121.8 &
     V364 Peg     &122.4 &
     DV UMa       &123.6 &
     V419 Lyr     &124.4 \\
     YZ Cnc       &125.0 &
     DM Dra       &125.3 & 
     V344 Lyr     &126.1 &
     GZ Cnc       &126.9 \\
     GX Cas       &128.2 &
     UV Gem       &128.2 \\
     \hline
    \end{tabular}
  \end{center}
\end{table*}


\end{document}